\documentclass[graybox]{svmult}

\usepackage{mathptmx}       
\usepackage{helvet}         
\usepackage{courier}        
\usepackage{amssymb}
\usepackage{makeidx}         
\usepackage{graphicx}        
\usepackage{multicol}        
\usepackage[bottom]{footmisc}

\makeindex             

\begin{document}

\title*{Quasinormal Modes Beyond Kerr}
\author{Aaron Zimmerman, Huan Yang, Zachary Mark, Yanbei Chen, Luis Lehner}
\institute{Aaron Zimmerman \at Canadian Institute for Theoretical Astrophysics, 60 St. George Street, Toronto ON Canada, \email{azimmer@cita.utoronto.ca}
}
%
%
\maketitle

\abstract*{
The quasinormal modes (QNMs) of a black hole spacetime are the free, decaying oscillations of the spacetime, and are well understood in the case of Kerr black holes. 
We discuss a method for computing the QNMs of spacetimes which are slightly deformed from Kerr. We mention two example applications: the parametric, turbulent instability of scalar fields on a background which includes a gravitational QNM, and the shifts to the QNM frequencies of Kerr when the black hole is weakly charged. 
This method may be of use in studies of black holes which are deformed by external fields or are solutions to alternative theories of gravity.
}

\abstract{
The quasinormal modes (QNMs) of a black hole spacetime are the free, decaying oscillations of the spacetime, and are well understood in the case of Kerr black holes. 
We discuss a method for computing the QNMs of spacetimes which are slightly deformed from Kerr. We mention two example applications: the parametric, turbulent instability of scalar fields on a background which includes a gravitational QNM, and the shifts to the QNM frequencies of Kerr when the black hole is weakly charged. 
This method may be of use in studies of black holes which are deformed by external fields or are solutions to alternative theories of gravity.
}

\section{Introduction}
\label{sec:Intro}

The quasinormal modes (QNMs) of a black hole are the characteristic oscillations of the spacetime. They decay in time as energy flows through the horizon and disperses to infinity as waves. 
They are present for all types of perturbations (scalar, electromagnetic, and gravitational), and are excited whenever the spacetime is transiently perturbed. 
In particular, the gravitational wave QNMs make up the ringdown phase following a binary black hole merger or the formation of a black hole, and so they play an important role in gravitational wave astrophysics. 
Reference~\cite{Berti2009} provides a detailed review of black hole QNMs.

The QNM frequencies are the solutions to an eigenvalue problem, analogous to the problem of computing the energy spectrum of the hydrogen atom in quantum mechanics. 
Efficient methods are available for computing the QNM frequencies and wavefunctions for Kerr black holes (e.g. Leaver's method~\cite{Leaver1985}), as well as many spherically symmetric generalizations of the Schwarzschild black hole. 
It is difficult, however, to solve for the frequency spectrum for generalizations of Kerr black holes, because the fundamental equations may no longer separate, or even decouple (as in the case of Kerr-Newman black holes).

Here we describe a new method for computing small shifts to the QNM frequencies for spacetimes which are slightly deformed from Kerr. 
This method is analogous to the problem of solving for the shifts in the energy levels of a quantum system that is perturbed by a small change in the Hamiltonian. 
We have applied this method to explore a parametric instability among the oscillation modes of rapidly rotating black holes~\cite{Yang2014}, and are currently using the method to calculate the QNM frequencies of weakly charged black holes with generic spin for the first time.

\section{Quasinormal Modes of Deformed Spactimes}
\label{sec:Perts}

Perturbations of rotating black holes are well understood, and can be treated using the Teukolsky formalism~\cite{Teukolsky1973} which unifies scalar, electromagnetic, and gravitational perturbations into a single master equation. 
This equation can be solved for spin-weighted scalar quantities ${}_s \psi$, where $s=0, \pm 1, \pm 2$ correspond to scalar, electromagnetic, and gravitational perturbations respectively. 
We write the master equation schematically as
\begin{equation}
\label{eq:Master}
L_T \left[ {}_s \psi(x^\mu) \right] = \mathcal T\,,
\end{equation}
where $L_T$ is a second order differential operator constructed using the Newman-Penrose formulation of the field equations~\cite{NewmanPenrose1962}, and $\mathcal T$ represents sources of stress energy which generate the perturbations. 
Since we are interested in the free oscillation frequencies of the spacetime, we set $\mathcal T =0$. All relevant gravitational and electromagnetic fields can be reconstructed from the spin-weighted scalars ${}_s \psi$. 
Remarkably, Eq.~\ref{eq:Master} separates when we expand the scalars in the frequency domain,
\begin{equation}
{}_s \psi = \sum_{lm} \int d \omega \,  \E^{- \I \omega t+ \I m \phi} {}_s S_{lm\omega} (\theta) {}_s R_{lm \omega} (r)\,,
\end{equation}
where ${}_s S_{lm\omega} $ are the spin-weighted spheroidal harmonics, and ${}_s R_{lm\omega}$ are the radial wavefunctions~\cite{Teukolsky1973}.
Inserting this expansion into Eq.~\ref{eq:Master} results in a pair of coupled eigenvalue equations for ${}_s S_{lm\omega}$ and ${}_s R_{lm \omega}$, which give the angular eigenvalues ${}_s A_{lm} (\omega)$ and the QNM frequencies $\omega_{lm}$.

We are interested in how this situation changes when we introduce a small deformation to the black hole spacetime, $g_{\mu \nu} = g^{(0)}_{\mu \nu} + \eta h^{(1)}_{\mu\nu}$ with $\eta \ll 1$. 
We call this a ``deformation'' to distinguish $h^{(1)}_{\mu \nu}$ from the further QNM perturbations of the spacetime.
The deformation can represent any small modification to the spacetime, for example the impact of additional multipole moments if the central object is a ``bumpy'' black hole, or the addition of a small amount of charge to the hole (which carries additional complications). 
As we proceed, we only keep terms linear in $\eta$. 

The small deformation should only change the QNM spectrum slightly, introducing frequency shifts $\omega \to \omega^{(0)} + \eta \omega^{(1)}$. 
To compute the frequency shifts $\omega^{(1)}$, we first write $\psi_{lm\omega} (r,\theta) = {}_s S_{lm \omega} \, {}_s R_{lm\omega}$ and expand the master equation in frequency and azimuthal harmonics,
\begin{equation}
\tilde L_T \left[ \psi_{lm\omega} (r,\theta) \right] = 0\,.
\end{equation}
Introducing the deformation and the shifts to the QNM frequencies gives
\begin{equation}
\label{eq:PertMaster}
\left (\tilde L_T^{(0)} + \eta \omega^{(1)} \frac{\partial \tilde L_T^{(0)}}{\partial \omega} + \eta \tilde L_T^{(1)} \right) \left[ \psi_{lm\omega}^{(0)} + \eta  \psi_{lm\omega}^{(1)} \right] \approx 0 \,.
\end{equation}
Equation~\ref{eq:PertMaster} is a perturbed eigenvalue equation, and is reminiscent of the problem of solving for the eigenvalues of a perturbed Hamiltonian in quantum mechanics. 
In that case, we have a Hamiltonian $H = H^{(0)} + \eta H^{(1)}$, perturbed energies $E_n \approx E^{(0)}_n + \eta E^{(1)}_n$, and perturbed eigenstates $| n \rangle \approx | n^{(0)} \rangle + \eta | n^{(1)} \rangle$. 
In order to solve for $E^{(1)}_n$, we left multiply the equation $H | n \rangle = E_n | n \rangle$ with background energy eigenstate $\langle n^{(0)}|$, expand to leading order in $\eta$, and arrive at
\begin{equation}
E_n^{(1)}  = \frac{\langle n^{(0)} | H^{(1)} | n^{(0)} \rangle}{\langle n^{(0)} | n^{(0)} \rangle} \,.
\end{equation}
Here we have written $E_n^{(1)}$ in a manner that emphasizes that this expression does not require a normalized (or even orthogonal) basis. 
We only require that $\langle n^{(0)} | (H^{(0)} - E^{(0)}_n) | n^{(1)}\rangle =0$, by acting the Hamiltonian on the left. 
In order to solve for the shifted energies, we need a self-adjoint, finite inner product $\langle | \rangle$. 

We can isolate the shifted frequencies of Eq.~\ref{eq:PertMaster} by defining such a product. However, at spatial infinity the outgoing boundary conditions for the radial functions mean that ${}_s R_{lm\omega} \sim r^{-2s -1} \E^{\I \omega_R r_* + \omega_I r_*}$ where $\omega = \omega_R - \I \omega_I$. 
In order for the QNMs to decay in time $\omega_I > 0$, so that the radial wavefunctions diverge at infinity. 
By analytically continuing ${}_sR_{lm\omega}$ into the complex-$r$ plane, these same solutions are exponentially decaying as we move into the upper half plane, $r \to + \I \infty$. Thus we define the finite inner product between two wavefunctions with the asymptotic behavior of QNMs,
\begin{equation}
\langle \psi | \chi \rangle = \int_{\mathcal C} (r-r_+)^s (r - r_-)^s dr \int \sin \theta d \theta \psi(r,\theta) \chi(r,\theta) \,,
\end{equation}
where the contour $\mathcal C$ begins at $+ i \infty$, encircles a branch cut in ${}_sR_{lm\omega}$ running from the horizon $r_+$ parallel to the imaginary axis, and returns to $+ \I \infty$ on the other side of the cut. The weights of the integral guarantee that the product is self-adjoint, so that $\langle \psi^{(0)}_{lm\omega} | \tilde L^{(0)}_T [\psi^{(1)}_{lm\omega}\rangle=0$. Using this inner product, the frequency shifts are
\begin{equation}
\omega^{(1)} = - \frac{\langle \psi^{(0)}_{lm\omega} | \tilde L_T^{(1)} | \psi^{(0)}_{lm\omega} \rangle}{\langle \psi^{(0)}_{lm\omega} | \partial_\omega \tilde L_T^{(0)} | \psi^{(0)}_{lm\omega} \rangle} \,.
\end{equation}
Note that this procedure works even when $\tilde L_T^{(1)}$ cannot be separated in $r$ and $\theta$. 
We have tested that this method recovers the corrections to the QNM frequencies of Schwarzschild when slow rotation is added.

\section{Parametric Instability of Rapidly Rotating Kerr Black Holes}
\label{sec:Param}

As a first application of the method described above, we have investigated the possibility of parametric resonances around rapidly rotating Kerr black holes. 
For rapidly rotating black holes, with $\epsilon = 1-a \ll 1$, the QNM spectrum is weakly damped and nearly evenly spaced in the azimuthal mode number $m$, $\omega = m/2 + O(\sqrt{\epsilon})$. 
This spacing leads to the possibility of resonant mode interactions beyond linear order in perturbations about Kerr.
In particular, parametric resonance is a possibility, since parametric resonance occurs when an oscillator is driven at a frequency $\omega$ approximately twice its natural frequency, $\omega \approx 2 \omega'$. 
This happens when a QNM of mode number $m$ drives a mode with half its mode number $m' = m/2$. The strength of the driving depends on the small amplitude of the driving mode, and competes with the slow $O(\sqrt{\epsilon})$ decay of the driven mode.

To investigate the possibility that an initially excited QNM can parametrically drive one or more modes into growth, we must use second order perturbation theory. 
Since this is quite challenging, we instead deal with the simpler but conceptually similar situation of a scalar field evolving in a dynamic background consisting of the black hole plus a weak graviational QNM perturbation. 
The QNM then serves as the driving mode, and the scalar field as a test oscillator that can be driven. 

The scalar field obeys
\begin{equation}
\label{eq:PertScalarWave}
\square_{g^{(0)} + h^{(1)}} \psi = \left ( \square_{g^{(0)}} + \mathcal{H}[h^{(1)}] \right ) [ \psi ] = 0,
\end{equation}
where $\mathcal H$ is the linearization of the wave operator with respect to $h^{(1)}_{\mu\nu}$.
Using the ansatz that the scalar field solution is a background QNM with an additional time dependence,
\begin{equation}
\psi \sim \E^{ \int \alpha(t)  dt - \omega_I' t} \E^{- \I (\omega_R/2) t + \I (m/2) \phi} \left(\psi^{(0)}_{lm\omega'} + \psi_{lm\omega'}^{(1)}  \right) + {\rm c.c.} \,,
\end{equation}
we can use the method described above to solve for the growth parameter $\alpha(t)$, which is time-dependent because the background QNM decays in time. 
In the above equation, c.c. indicates the complex conjugate of the preceding terms.
Our self-adjoint inner product allows us to eliminate the unknown correction to the wavefunctions $\psi_{lm\omega}^{(1)}$, and $\alpha(t)$ can be determined solely from integrals involving the unperturbed QNM wavefunctions and the metric.
Whether or not growth occurs for some time is a sensitive function of the initial amplitude of the gravitational perturbation, but may be possible for a moderately small mass ratio merger, provided the final black hole is rapidly spinning~\cite{Yang2014}.
The growth is always transient, because the driving mode decays in time.

An intriguing aspect of this instability is the fact that a single gravitational mode with azimuthal number $m$ can drive modes with many different angular numbers $l$, provided they have azimuthal number $m' = m/2$. 
In addition, modes of lower frequencies are driven by modes with higher frequencies. This situation is analogous to the recently discovered turbulent behavior of perturbations of 4-dimensional AdS black hole spacetimes~\cite{Adams:2013vsa} and their 3-dimensional conformal fluid duals~\cite{Green:2013zba}, which feature an inverse cascade of mode energy to lower frequencies. 
In fact, it is the connection to the fluid-gravity correspondence that first motivated the study of the parametric instability, which may represent the onset of turbulence around a rapidly rotating black hole~\cite{Yang2014}. 
This would be the first example of turbulent gravitational behavior in an asymptotically flat spacetime.

\section{The Quasinormal Modes of Kerr-Newman Black Holes}
\label{sec:KN}

A second application of this method is to weakly charged, rotating black hole spacetimes. 
These are the small-charge limit of Kerr-Newman (KN) black holes, parameterized by both $a$ and the charge $Q$ in geometric units.
In this case, the background spacetime contains electromagnetic (EM) fields. 
A perturbation to the gravitational fields couples to the background EM fields and vice versa, so that these types of perturbations are coupled. Because of this, the problem of perturbations of KN black holes has only been treated in the cases $a=0$ (Reissner Nordstrom black holes)~\cite{Leaver1990} and $a \ll 1$~\cite{Pani:2013ija,Pani:2013wsa}.
The methods discussed here allow for the computation of the QNM frequencies in small charge limit, where $q = Q^2 \ll 1$. 
In this case the KN solution is simply a small deformation of Kerr.

The perturbation equations were derived by Chandrasekhar in the phantom gauge for the coupled $s=2$ and $s=1$ scalars~\cite{ChandraBook}, and are schematically given by
\begin{eqnarray}
\label{eq:GravKN}
\left[\tilde L_T^{(0)} + q \tilde L_T^{(1)} + q\omega^{(1)} \partial_\omega \tilde L_T^{(0)}\right] [{}_2\psi_{lm\omega}] = q \tilde G^{(1)}[{}_1\psi_{lm\omega}] \,, \\
\label{eq:EMKN}
\left[\tilde L_T^{(0)} + q \tilde L_T^{(1)} + q\omega^{(1)} \partial_\omega \tilde L_T^{(0)}\right] [{}_1\psi_{lm\omega}] = q \tilde F^{(1)}[{}_2\psi_{lm\omega}] \,.
\end{eqnarray}
However, we are interested in how the background gravitational  $(s=2)$ modes change with the addition of a small charge, and in this case the electromagnetic  $(s=1)$ modes must be $O(q)$,
\begin{eqnarray}
{}_2\psi_{lm\omega} & = & {}_2\psi_{lm\omega}^{(0)} + q \; {}_2\psi_{lm\omega}^{(1)} + O(q^2) \,, \\
{}_1\psi_{lm\omega} & = & q \; {}_1\psi_{lm\omega}^{(1)} + O(q^2) \,.
\end{eqnarray}
Inserting this expansion into the pair of Eqs.~\ref{eq:GravKN} and~\ref{eq:EMKN}, we recover Eq.~\ref{eq:PertMaster} for the shift of the gravitational QNM frequencies. Similarly, we can recover the leading correction to the electromagnetic QNM frequencies.

\section{Outlook}
\label{sec:Outlook}

The eigenvalue perturbation method described here can be applied to a variety of situations of theoretical and astrophysical interest. 
It can be applied directly when the deformation of the black hole spacetime is stationary and axisymmetric, and it can also be generalized to dynamical or non-axisymmetric deformations when these are decomposed into modes, as the case of the parametric instability illustrates.
In addition, the analysis of the Kerr-Newman black hole indicates that the method applies even when coupling exists between perturbations and the sources of the deformed metric, such as stress-energy in the spacetime.
Such couplings shift the QNM frequencies beyond $O(\eta)$ and can be neglected at that order.
The method is quite general, and can be applied to deformations of other systems whose QNM spectrum is known. 
In particular, application to higher dimensional black holes may provide new insights about these objects.

A major technical challenge is to compute the correction to the Teukolsky equation, $\tilde L^{(1)}_T$ given the deformation $h^{(1)}_{\mu \nu}$ to the background metric, but in principle this is straightforward. 
The case of scalar fields is especially simple. A general procedure for generating $\tilde L^{(1)}_T$ for all spin-weights is left to future work, along with additional applications of the eigenvalue perturbation method.

We thank Stephen Green and Rob Owen for useful conversations during the course of this work. 
AZ, HY, and YC were supported by NSF Grant PHY-1068881, CAREER Grant 0956189, and the David and Barbara Groce Startup Fund at Caltech.  
ZM was supported by the LIGO SURF program at Caltech.
LL was supported by NSERC through Discovery Grants and CIFAR. 
This research was supported in part by Perimeter Institute for Theoretical Physics. 
Research at Perimeter Institute is supported by the Government of Canada through Industry Canada and by the Province of Ontario through the Ministry of Research and Innovation.

\end{document}